\documentclass[spanish,english,aps,amssymb,preprint]{revtex4-1}
\usepackage[latin9]{inputenc}
\usepackage{color}
\usepackage{amsmath}
\usepackage{graphicx}

\makeatletter

\providecommand{\tabularnewline}{\\}

\DeclareMathOperator{\sn}{sn}
\DeclareMathOperator{\cn}{cn}
\DeclareMathOperator{\dn}{dn}

\DeclareMathOperator{\re}{Re}
\DeclareMathOperator{\im}{Im}

\usepackage{babel}

\begin{document}

\title{The Elliptic Gaudin Model: a Numerical Study}

\author{C. Esebbag}

\affiliation{Departamento de F\'{i}sica y Matemáticas. Universidad de Alcalá.
28871 Alcalá de Henares, Spain}

\author{J. Dukelsky}

\affiliation{Instituto de Estructura de la Materia, CSIC, Serrano 123, E-28006
Madrid, Spain}
\begin{abstract}
The elliptic Gaudin model describes completely anisotropic spin systems
with long range interactions. The model was proven to be quantum integrable
by Gaudin and latter the exact solution was found by means of the
algebraic Bethe ansatz. In spite of the appealing properties of the
model, it has not yet been applied to any physical problem. We here
generalize the exact solution to systems with arbitrary spins, and
study numerically the behavior of the Bethe roots for a system with
three different spins. Then, we propose an integrable anisotropic
central spin model that we study numerically for very large systems.
\end{abstract}
\maketitle

\section{introduction}

In 1976 Michel Gaudin proposed three quantum integrable models for
spin 1/2 chains with infinite range interactions \cite{Gaudin}. Two
of these models, the rational or XXX and the hyperbolic-trigonometric
or XXZ models, were solved for the spectrum and eigenstates. However,
the exact solution of the third integrable model, the Elliptic Gaudin
Model (EGM) or XYZ model, had to wait till 1996 \cite{Sklyanin} for
a complete solution in terms of the Algebraic Bethe Ansatz (ABA).
The Gaudin models played an important role in the development and
testing of several methods in quantum integrable theory, like the
functional Bethe ansatz and the separation of variables \cite{Skl2,Skl3},
the relation with the Knizhnik-Zamolodchikov equations \cite{Bab,babujian}
and with the corresponding Wess-Zumino-Witten models \cite{WZW},
the construction of Bäcklund transformations \cite{Zullo}, etc. On
a different perspective, the rational Gaudin model \cite{Amico,Duk1}
was linked to the exact solution of the reduced Bardeen-Cooper-Schriefer
(BCS) Hamiltonian solved exactly by Richardson \cite{Rich1} and proved to be quantum integrable by Cambiaggio, Rivas and Saraceno \cite{CRS}  .
Exploiting this link, several families of exactly solvable models
called Richardson-Gaudin (RG) models were proposed \cite{Amico,Duk1}.
Since then, the rational RG model found important applications to
different areas of quantum many-body physics including ultrasmall
superconducting grains \cite{Delft,Links1}, Tavis-Cummings models
\cite{atom-mol,Lerma}, cold atomic gases \cite{Schuck,Links2,BCS-BEC},
quantum dots \cite{Bortz,Fari} and nuclear structure \cite{Cooper,Heavy}
(for a review see \cite{Duk2,Ortiz}). More recently, the hyperbolic
RG model was applied to describe p-wave pairing in 2D lattices \cite{Sierra,Romb,Stin}
and 1D Kitaev wires \cite{Kitaev}. With less success, there have
been attempts to apply the EGM to matter-radiation problems including
counter-rotating terms \cite{Kundu,Hikami}. However, these integrable
models lack of the radiative excitation term or lead to non-hermitian
Hamiltonians. On a different respect, the EGM has been used to study
the thermalization process of a spin chain with long range interactions
in the transition from integrability to chaos \cite{Relano}. None
of these works attempted to find a numerical solution of the Bethe
equations.

The aim of this paper is to survey and generalize the exact solution
of EGM for arbitrary SU(2) spins systems, and to study numerically
the properties of the exact solutions in different scenarios. We first
introduce the model with the corresponding exact solution in Sec.
II. We then discuss a simple model of three different spins in Sec.
III for which we show how to solve the Bethe equations in order to
obtain the complete set of eigenstates. Next, we move to a physically
oriented problem, a new integrable anisotropic central spin model (ACSM), in Sec. IV. In Sec. V we solve exactly the ACSM for long
chains, and we extrapolate the ground state energy to the large $N$
limit, showing that it coincides with the classical spin approximation
in the thermodynamic limit.

\section{The Elliptic Gaudin Model}

The EGM was first proposed by Gaudin\cite{Gaudin} as a particular
family of integrable spin Hamiltonians with a fully anisotropic spin-spin
interactions. The $N$ commuting integrals of motion for a system
of $N$ arbitrary spins $S_{i}^{\alpha}$, with $\alpha=x,y,z$ and
$i=1,\cdots,N$ are
\begin{equation}
R_{i}=\sum_{{j=1\atop (j\neq i)}}^{N}J_{ij}^{x}S_{i}^{x}S_{j}^{x}+J_{ij}^{y}S_{i}^{y}S_{j}^{y}+J_{ij}^{z}S_{i}^{z}S_{j}^{z},\label{Rxyz}
\end{equation}
where the matrices $J_{ij}^{\alpha}$ satisfy the Bethe equations
\begin{eqnarray*}
J_{ij}^{\alpha}J_{jk}^{\gamma}+J_{ji}^{\beta}J_{ik}^{\gamma}+J_{ik}^{\alpha}J_{kj}^{\beta} & = & 0,
\end{eqnarray*}
in order to fulfill the integrability conditions $\left[R_{i},R_{j}\right]$=0.

Following Ref. \cite{babujian}, the $J_{ij}^{\alpha}$ can be expressed
in terms of the doubly periodic elliptic Jacobi functions of modulus
$k$, $\sn(z,k)$, $\cn(z,k)$ and $\dn(z,k)$ (for brevity, in general,
we will not explicitly write the modulus), and a set of $N$ arbitrary
coefficients $z_{i}$ as

\[
J_{ij}^{x}=\frac{1+k~\sn^{2}(z_{i}-z_{j})}{\sn(z_{i}-z_{j})},
\]

\begin{equation}
J_{ij}^{y}=\frac{1-k~\sn^{2}(z_{i}-z_{j})}{\sn(z_{i}-z_{j})}\label{Jotas},
\end{equation}

\[
J_{ij}^{z}=\frac{~\cn(z_{i}-z_{j})\,\dn(z_{i}-z_{j})}{\sn(z_{i}-z_{j})}.
\]

Alternatively, the integrals of motion $R_{i}$ can be expressed in
terms of the raising and lowering spin operators $S^{\pm}=S^{x}\pm iS^{y}$
as

\begin{align}
R_{i} & =\sum_{j\left(\neq i\right)}\left[\frac{k}{2}~\sn(z_{i}-z_{j})~\left(S_{i}^{+}S_{j}^{+}+S_{i}^{-}S_{j}^{-}\right)+\frac{1}{2\sn(z_{i}-z_{j})}\left(S_{i}^{+}S_{j}^{-}+S_{i}^{-}S_{j}^{+}\right)\right.\nonumber \\
 & \left.+\frac{~\cn(z_{i}-z_{j})\dn(z_{i}-z_{j})}{\sn(z_{i}-z_{j})}S_{i}^{z}S_{j}^{z}\right].\label{RSS}
\end{align}

The elliptic integrals of motion break the $su(2)$ symmetry associated
with the conservation of the total spin $S$ and the $u(1)$ symmetry
associated with the conservation of the $z$ component of the total
spin $S^{z}$. However, the model preserves a discrete $Z_{2}$ symmetry
associated with a $\pi$ rotation of every spin around an arbitrary
axis. Assuming $z$ as the quantization axis, a rotation by an angle
$\pi$ around this axis is related to the parity operator $P=\prod_{i=1}^{N}e^{i\pi\left(S_{i}^{z}+s_{i}\right)}$
with eigenvalues $+1$ for positive parity and $-1$ for negative
parity. Therefore, the complete set of common eigenstates of the integrals
of motion can be classified in two subsets of even or odd parity.

The exact solution comprising the eigenvalues of the integrals of
motion and the equations for determining the spectral parameters for
a system of $N$ spins with $s_{i}=\frac{1}{2}$ has been obtained
by means of the ABA in references \cite{Sklyanin,babujian}. Here, we
 present the exact solution for a general system of $N$ arbitrary
spins $s_{i}$. The derivation starting from a system of 1/2 spins is given in the Appendix.
The eigenvalues $r_{i}$ of the integrals of motion (\ref{Rxyz}), depending
on a set of $M$ roots $\lambda_{\alpha}$ to be determined by the
Bethe equations that will be introduced below, are:
\begin{equation}
r_{i}=s_{i}\left[\sum_{j(\neq i)}^{N}\,s_{j}\,\left(\varphi_{4}(z_{i}-z_{j})+\varphi_{1}(z_{i}-z_{j})\right)-\sum_{\alpha=1}^{M}\left(\varphi_{4}(z_{i}-\lambda_{\alpha})+\varphi_{1}(z_{i}-\lambda_{\alpha})\right)+i\frac{\pi l}{2K}\right],\label{autov}
\end{equation}
where $M=\sum_{i=1}^{N}s_{i}$. Any combination of spins  $s_{i}$
is allowed as long as the resulting summation $M$ is integer. Notice
that $M=N/2$ for the $s_{i}=1/2$ spin case and therefore, it would only
admit an exact solution for an even number of spins $N$. In addition,
$l$ is an integer number that can take the values 0 or 1 in order
to identify the two parity sectors.

At this point, we have to introduce the Jacobi Theta functions \cite{Abramowitz,NIST}
$\Theta(u,k)=\vartheta_{4}\left(v|q\right)$ and $\mathrm{H}(u,k)=\vartheta_{1}\left(v|q\right)$,
with the variable transformation $v=\frac{\pi u}{2K}$. In these definitions
$K(k)=\int_{0}^{\pi/2}\frac{d\theta}{\sqrt{1-k^{2}\sin^{2}\theta}}$
is the complete elliptic integral of the first kind, $K^{\prime}(k)=K\left(\sqrt{1-k^{2}}\right)$
and the nome $q=e^{-\pi K^{\prime}/K}$. The functions $\varphi_{1}(u)$,
and $\varphi_{4}(u)$ can be defined now following \cite{Gould},
as :
\[
\varphi_{1}(u)=\frac{H'(u)}{H(u)}\,,\qquad\qquad\varphi_{4}=\frac{\Theta'(u)}{\Theta(u)}\;.
\]

The $M$ roots $\lambda_{\alpha}$ in eq (\ref{autov}) are determined
by solving the set of $M$ coupled Bethe equations (see the Appendix):

\begin{flalign}
\sum_{j=1}^{N}\,s_{j}\,\left(\varphi_{4}(\lambda_{\alpha}-z_{j})+\varphi_{1}(\lambda_{\alpha}-z_{j})\right)-\sum_{\beta(\neq\alpha)}^{M}\left(\varphi_{4}(\lambda_{\alpha}-\lambda_{\beta})+\varphi_{1}(\lambda_{\alpha}-\lambda_{\beta})\right)+i\frac{\pi l}{2K} & =0\,.\label{eq:Equ}
\end{flalign}

The function $\varphi(\lambda)=\varphi_{1}(\lambda)+\varphi_{4}(\lambda)$
appearing in the eigenvalues (\ref{autov}) and in the Bethe equations
(\ref{eq:Equ}) has the special property of being periodic in the
real part of its argument but ``quasi periodic'' in the imaginary
part,\textit{ i.e.} $\varphi(\lambda+i\,K^{\prime})=\varphi(\lambda)+i\,C(k)$,
where $K^{\prime}$ is the quasi-period (in the imaginary direction)
and $C(k)$ is a real constant only depending on the elliptic modulus
$k$. As a consequence the imaginary part of the Bethe roots are constrained
to its natural interval $\mathrm{Im}(\lambda)\in[-\frac{K^{\prime}}{2},\frac{K^{\prime}}{2}]$.
Numerical solutions obtained outside of this interval may lead to
spurious non-physical results. On the other hand, as the real period
of $\varphi(\lambda)$ is $2K$, the Bethe roots of any physical solution
should be constraint to the fundamental rectangle of the complex plain
given by $[0,2K]\times[-\frac{K^{\prime}}{2},\frac{K^{\prime}}{2}]$
.

We next analyze the behavior of the integrals $R_{i}$ in the two
limits $k\to0$ and $k\to1$. Taking into account that for $k\to0$,
$\sn\left(u,k\right)\to\sin\left(u\right)$, $\cn\left(u,k\right)\to\cos\left(u\right)$
and $\dn\left(u,k\right)\to1$, it is easy to check that the elliptic
integrals $R_{i}$ (\ref{RSS}) transform into the trigonometric ones:

\[
R_{i}^{0}=\sum_{j\left(\neq i\right)}\left[\frac{1}{2\sin(z_{i}-z_{j})}\left(S_{i}^{+}S_{j}^{-}+S_{i}^{-}S_{j}^{+}\right)+\cot(z_{i}-z_{j})\,S_{i}^{z}S_{j}^{z}\right].
\]

On the other hand, starting from the $x,y,z$ representation (\ref{Rxyz})
and taking the limit $k\to1$, the elliptic functions transform according
to $\sn\left(u,k\right)\to\tanh\left(u\right)$ and $\cn\left(u,m\right),\,\dn(u,k)\,\to\:\mathrm{sech}\left(u\right)$,
thus we obtain:

\[
R_{i}^{1}=\sum_{j\left(\neq i\right)}\frac{1}{2\cosh(z_{i}-z_{j})\sinh(z_{i}-z_{j})}\left(S_{i}^{y}S_{j}^{y}+S_{i}^{z}S_{j}^{z}\right)+\frac{\cosh^{2}(z_{i}-z_{j})+\sinh^{2}(z_{i}-z_{j})}{\cosh(z_{i}-z_{j})\sinh(z_{i}-z_{j})}S_{i}^{x}S_{j}^{x}.
\]

Performing a cyclic permutation of the axis and making use of some
identities of the hyperbolic functions, we finally obtain the Gaudin
hyperbolic integrals:

\[
R_{i}^{1}=\sum_{j\left(\neq i\right)}\left[\frac{1}{2\sinh(\eta_{i}-\eta_{j})}\left(S_{i}^{+}S_{j}^{-}+S_{i}^{-}S_{j}^{+}\right)+\coth(\eta_{i}-\eta_{j})\,S_{i}^{z}S_{j}^{z}\right],
\]

where $\eta_{i}=2\,z_{i}$.

In a similar way, it can be shown that the eigenvalues (\ref{autov})
and Bethe equations (\ref{eq:Equ}), reduce to the corresponding
trigonometric and hyperbolic Gaudin solutions.

\section{A three-spin system}

In order to understand the behavior of the Bethe roots for the different
eigenstates we treat in this section the simplest integrable EGM with
three different spins, $N=3$, and $s_{1}=1/2$, $s_{2}=1$, $s_{3}=3/2$.
We construct an exactly solvable spin Hamiltonian as a linear combination
of the integrals of motion

\begin{equation}
H=\sum_{i<j}^{3}\sum_{\alpha=1}^{3}H_{ij}^{\alpha}S_{i}^{\alpha}S_{j}^{\alpha}\,,\label{ham3}
\end{equation}
We choose for the Hamiltonian $H=-\frac{1}{2}R_{1}-\frac{1}{4}R_{2}$,
with the parameters $z_{i}=0,\,0.2,\,0.4$ and the elliptic modulus
$k=\frac{1}{2}$. The corresponding coefficients $H_{ij}^{\alpha}$
are displayed in Table \ref{tab:coef}.

\begin{table}
\protect %
\begin{tabular}{|c|c|c|c|c|c|c|c|c|}
\hline
$H_{12}^{x}$  & $H_{12}^{y}$  & $H_{12}^{z}$  & $H_{13}^{x}$  & $H_{13}^{y}$  & $H_{13}^{z}$  & $H_{23}^{x}$  & $H_{23}^{y}$  & $H_{23}^{z}$\tabularnewline
\hline
1.28522  & 1.23563  & 1.2293  & 1.38861  & 1.19509  & 1.16865  & 1.28522  & 1.23563  & 1.2293\tabularnewline
\hline
\end{tabular}\protect\caption{\label{tab:coef} Exchange couplings of the three-spin Hamiltonian
(\ref{ham3}).}
\end{table}

The dimension of the Hilbert space is $d=\prod_{i=1}^{3}\left(2s_{i}+1\right)=24$.
The Hamiltonian matrix is block diagonal with $d_{+}=12$ for positive
parity and $d_{-}=12$ for negative parity. The number of Bethe equations
(\ref{eq:Equ}) and roots is $M=\sum_{i=1}^{3}s_{i}=3$. For each
parity sector, defined by $l=0$ or $1$, we look for 12 different
solutions with the three roots restricted to the fundamental rectangle
defined by $0<\mathrm{Re}(\lambda)<2K$ and $-\frac{K^{\prime}}{2}\leq\mathrm{Im}(\lambda)\leq\frac{K^{\prime}}{2}$.
For this particular case $2K=3.3715$ and $\frac{K^{\prime}}{2}=1.07826.$

Tables \ref{positive} and \ref{negative} depict the complete set
of eigenvalues and the corresponding values of the Bethe roots for
positive and negative parity respectively. For positive parity $\left(l=0\right)$
and real parameters $z_{i}$, the Bethe roots are real or complex
conjugate pairs with the exception of roots having the imaginary part
equal to half the imaginary quasi-period $\left(\pm\frac{K'}{2}\right)$.
In the latter case the pair of complex roots need not to be a conjugate
pair (real parts could be different). Moreover, complex conjugation
of all roots gives the same solution due to the quasi-periodicity
of the Jacobian functions. On the other hand, for negative parity
$\left(l=1\right)$ one of the roots is always complex with the imaginary
part equal to half the imaginary quasi-period $\left(+\frac{K'}{2}\right)$,
to compensate the imaginary term added to the Bethe equations. The
other roots could be real, or complex pairs following the same behavior
as in the $l=0$ case.

As a final remark, we have checked that the eigenvalues obtained solving
the Bethe equations (\ref{eq:Equ}) fully agree with the results of
an exact diagonalization of the Hamiltonian $\left(\ref{ham3}\right)$.
\begin{table}
\begin{tabular}{|c|c|c|c|}
\hline
$E$  & $\lambda_{1}$  & $\lambda_{2}$  & $\lambda_{3}$ \tabularnewline
\hline
-8.13147  & 0.277673  & 0.261164 + 0.115827 \textit{i}  & 0.261164 - 0.115827 \textit{i} \tabularnewline
\hline
-5.64950  & 0.0674548  & 0.268200  & 2.150100 \tabularnewline
\hline
-5.48168  & 1.92346  & 0.281146 + 0.0491301 \textit{i}  & 0.281146 -- 0.0491301 \textit{i} \tabularnewline
\hline
-0.850805  & 0.286691  & 0.726263 + 1.07826 \textit{i}  & 3.15855 -- 1.07826 \textit{i} \tabularnewline
\hline
-0.758290  & 0.286760 & 0.257994 + 1.07826 \textit{i}  & 1.94100 -- 1.07826 \textit{i} \tabularnewline
\hline
-0.649792 & 0.0448926  & 2.06330 + 0.519302 \textit{i}  & 2.06330\, -- 0.519302 \textit{i} \tabularnewline
\hline
-0.615993  & 0.0473203  & 0.375176 + 1.07826 \textit{i}  & 2.06325 -- 1.07826 \textit{i} \tabularnewline
\hline
-0.606659  & 0.0481756  & 0.825910 -- 1.07826 \textit{i}  & 3.29741 + 1.07826 \textit{i} \tabularnewline
\hline
-0.394121  & 0.287018  & 1.94224 + 0.516064 \textit{i}  & 1.94224 -- 0.516064 \textit{i} \tabularnewline
\hline
6.69616  & 1.95117  & 0.612035 -- 1.07826 \textit{i}  & 3.29405 + 1.07826 \textit{i} \tabularnewline
\hline
6.88050  & 1.95157  & 0.268084 -- 1.07826 \textit{i}  & 1.95185 + 1.07826 \textit{i} \tabularnewline
\hline
7.22436  & 1.95228  & 1.95248 + 0.704049 \textit{i}  & 1.95248 -- 0.704049 \textit{i} \tabularnewline
\hline
\end{tabular}\protect\caption{Eigenvalues and Bethe roots in the positive parity sector $(l=0)$
of the three-spin Hamiltonian (\ref{ham3}).}
\label{positive}
\end{table}

\begin{table}
\begin{tabular}{|c|c|c|c|}
\hline
$E$  & $\lambda_{1}$  & $\lambda_{2}$  & $\lambda_{3}$\tabularnewline
\hline
-5.86850  & 0.239016 + 1.07826 \textit{i}  & 0.280492 + 0.0487235 \textit{i}  & 0.280492 -- 0.0487235 \textit{i} \tabularnewline
\hline
-5.64331  & 0.0682364  & 0.268316  & 2.14920 + 1.07826 \textit{i} \tabularnewline
\hline
-5.59109  & 0.0727121  & 0.269232  & 0.458056 + 1.07826 \textit{i} \tabularnewline
\hline
-5.52799  & 0.281070 + 0.0490841 \textit{i}  & 0.281070 -- 0.0490841 \textit{i}  & 1.92361 + 1.07826 \textit{i} \tabularnewline
\hline
-0.714459  & 0.286793  & 1.94080  & 0.258161 + 1.07826 \textit{i} \tabularnewline
\hline
-0.649689  & 0.0448967  & 2.06348  & 2.06312 + 1.07826 \textit{i} \tabularnewline
\hline
-0.619842  & 0.0469765  & 2.06361  & 0.375167 + 1.07826 \textit{i} \tabularnewline
\hline
-0.395533  & 0.287017  & 1.94206  & 1.94243 + 1.07826 \textit{i} \tabularnewline
\hline
6.59546  & 0.267512 + 1.07826 \textit{i}  & 0.985596 + 1.07826  & 2.91839 -- 1.07826 \textit{i} \tabularnewline
\hline
6.64346  & 0.599786 + 1.07826 \textit{i} & 1.95133 -- 1.07826 \textit{i}  & 3.30613 + 1.07826 \textit{i} \tabularnewline
\hline
6.88448  & 0.268092 + 1.07826 \textit{i}  & 1.95170 + 0.496275 \textit{i}  & 1.95170 -- 0.496275 \textit{i} \tabularnewline
\hline
7.22431  & 1.95235 + 0.345399 \textit{i}  & 1.95235 -- 0.345399 \textit{i}  & 1.95255 +.07826 \textit{i} \tabularnewline
\hline
\end{tabular}\protect\caption{Eigenvalues and Bethe roots in the negative parity sector $(l=1)$
of the three-spin Hamiltonian (\ref{ham3}).}
\label{negative}
\end{table}

\section{\label{sec:CSM}The Anisotropic Central Spin Model}

\begin{figure}
\begin{centering}
\includegraphics[scale=0.6]{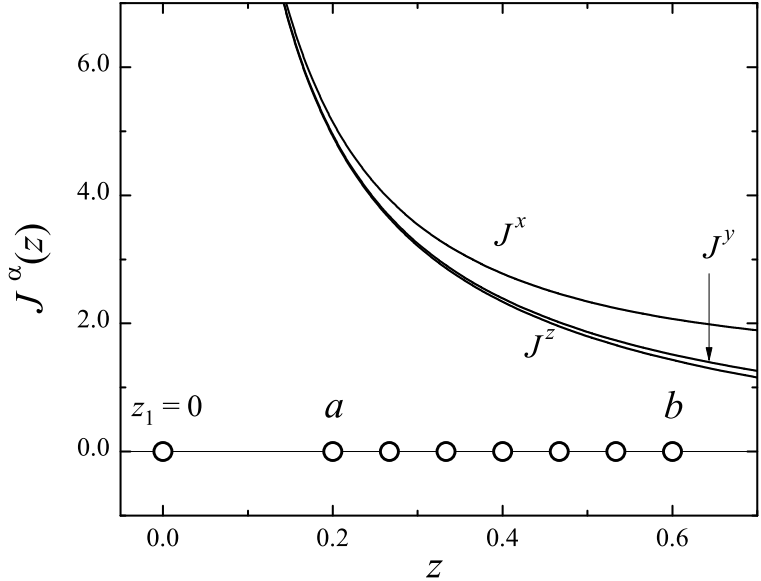}
\par\end{centering}

\centering{}\protect\protect\caption{\label{fig:Interaction} Exchange couplings of central spin Hamiltonian
(\ref{HCS}) as a function of distance. In the horizontal axis we
show as an example a system with $N=8$ spins.}
\end{figure}

The central spin model (CSM), describing the hyperfine interaction of
an electron spin in a quantum dot with a non-interacting system of
surrounding nuclear spins, has been proposed as the main component of spintronic devises and solid state qubits \cite{CSM-Rev}. The isotropic CSM Hamiltonian
with Heisenberg exchange couplings between the central spin and the
nuclear spin bath, subjected to an external magnetic field, is
precisely one of the integrals of motion of the rational RG
model. As such, it has been extensively studied using exact solutions
\cite{Bortz,Fari}. Anisotropic effects due to the hyperfine interaction between the central spin and the nuclear bath can still be investigated within the Hyperbolic or XXZ RG model \cite{Fisher}. However, the inclusion of the quadrupole coupling in the electron-bath interaction goes beyond the XXZ model \cite{Sini}.
Here, as a physically oriented example of a quantum integrable system
derived from the EGM, we study a modified anisotropic central spin
model (ACSM) without an external magnetic field. The introduction of an external magnetic field breaks the integrability of the EGM since it does not admit linear terms as opposed to the rational and trigonometric-hyperbolic cases. In our model the system is composed
by a single electronic spin $s_{1}=\frac{1}{2}$ interacting with
the $N-1$ nuclear spins $\frac{1}{2}$ of the bath. The hyperfine and quadrupole couplings between the electron and the
surrounding spins is described by a completely anisotropic antiferromagnetic
exchange interaction. We assume that the electron spin is located at position
$z_{1}=0$, while the environmental spins are uniformly distributed
within a linear segment at a finite distance $a$ $\left(z_{2}=a\right)$
with the last spin at position $z_{N}=b$, with $0<a<b\leq K$.
The restriction for $b$ to be lower than or equal to $K$ is necessary
to keep the interaction decreasing with distance in the selected interval.
Therefore, the values of the fixed parameters $z$ are given by $z_{i}=a+\frac{i-2}{N-2}(b-a)$
for $i>1$. The anisotropic central spin Hamiltonian is defined by
the first integral of motion (\ref{Rxyz}) of the EGM:

\begin{equation}
H_{ACSM}=-R_{1}=\sum_{j=2}^{N}\left(J_{j}^{x}S_{1}^{x}S_{j}^{x}+J_{j}^{y}S_{1}^{y}S_{j}^{y}+J_{j}^{z}S_{1}^{z}S_{j}^{z}\right)\,,\label{HCS}
\end{equation}
where $J_{j}^{\alpha}=-J_{1j}^{\alpha}(-z_{j})=J_{1j}^{\alpha}(z_{j})$
for $\alpha\equiv x,y,z$ as given in Eq. (\ref{Jotas}). The properties
of elliptic functions determine the relation between the exchange
interactions in the $x$, $y$, $z$ directions as $J_{j}^{x}>J_{j}^{y}>J_{j}^{z}$
~for~ $0<z_{j}<K$. Figure \ref{fig:Interaction} shows the three
components of the interaction as a function of distance for $k=0.5$.
In the horizontal axis we display, as an example, a spin network with
$N-1=7$ environmental spins uniformly distributed in a segment with
$a=0.2$ and $b=0.6$.

In order to gain insight into the structure of the GS wavefunction
we explore the classical description of the model. In this approximation each spin
$\frac{1}{2}$ operator is replaced by $S\equiv\frac{1}{2}[cos\varphi\,sin\theta,sin\varphi\,sin\theta,cos\theta]$,
where $\theta$ and $\text{\ensuremath{\varphi}}$ are the usual polar
and azimuthal angles satisfying $0\leq\theta\leq\pi$ and $0\leq\varphi\leq2\pi$.
Inserting into the Hamiltonian (\ref{HCS}), the classical energy
is given by
\begin{equation}
E_{cl}=\frac{1}{4}\sum_{j=2}^{N}\left(sin\theta_{1}\,sin\theta_{j}\left[J_{j}^{x}\,cos\varphi_{1}\,cos\varphi_{j}\,+J_{j}^{y}\,sin\varphi_{1}\,sin\varphi_{j}\right]+J_{j}^{z}\,cos\theta_{1}\,cos\theta_{j}\right)=\frac{1}{4}\,\sum_{j=2}^{N}\,E_{j}\,.\label{enerclas-1}
\end{equation}
Our task is to find the absolute minimum of the classical energy $E_{cl}$
with respect to the angular variables of the central spin $\theta_{1}$,
$\varphi_{1}$, and those of the nuclear spins $\{\theta_{j},\varphi_{j}\}_{j=2}^{N}$.
Each term of the sum $E_{j}$ depends exclusively on the angles of
a particular nuclear spin $j$ and the angles of the central spin.
Minimization with respect to the angles leads to a system of $2\times N$
coupled variational equations. However, minimizing each term $E_{j}$
independently will yield the absolute minimum if each solution is compatible with a unique value of
the central spin variables $\theta_{1},\varphi_{1}$.
Notice that even though the spins in the bath are non-interacting, the latter condition induces an effective interaction through the central spin. By solving
the four variational equations derived from the minimization of each
$E_{j}$ we obtain different classes of solutions corresponding to
stationary values of $E_{j}$ in the set $\{\pm J_{j}^{x},\,\pm J_{j}^{y},\,\pm J_{j}^{z}\}$.
As the largest component of the interaction is $J_{j}^{x}$ we conclude
that the minimum for each term is $E_{j}=-J_{j}^{x}$, and the corresponding
angles are $\theta_{1}=\frac{\pi}{2}$, $\varphi_{1}=\pi$, $\theta_{j}=\frac{\pi}{2}$,
$\varphi_{j}=0$ ($j>1)$. The classical GS corresponds to an antiferromagnetic
state with all spins aligned into the $x$ axis, and the central
spin pointing in the opposite direction to the environmental spins.
Therefore, for the minimal energy configuration,
the classical energy per spin is:

\begin{equation}
\frac{E_{cl}}{N}=-\frac{1}{4\,N}\,\sum_{j=2}^{N}\,J_{j}^{x}=-\frac{1}{4\,N}\,\sum_{j=2}^{N}\,\frac{1+k~\sn^{2}(z_{j})}{\sn(z_{j})}\;.\label{enerclas2-1}
\end{equation}

In order to find an expression for the classical energy
density in the thermodynamic limit, i.e. $\lim_{N\to\infty}\frac{E_{cl}}{N}$,
we define a normalized density distribution for the spins such that
$\intop_{\Omega}\rho(z)\,dz=1$, where $\Omega\subset(0,K)$ is the
compact interval containing all environmental spins (the parameters
$z$'s except for $z_{1}$), so that the number of spins in an interval
of length $dz$ is given by $(N-1)\,\rho(z)\,dz$. Introducing this
distribution in equation (\ref{enerclas2-1}) and taking the corresponding limit
we obtain in general:

\begin{equation}
\mathcal{E}_{cl}=\lim_{N\to\infty}\,\frac{E_{cl}}{N}=-\frac{1}{4}\,\intop_{\Omega}\,\frac{1+k~\sn^{2}(z)}{\sn(z)}\,\rho(z)\,dz\,.\label{classcont}
\end{equation}

According to our model of equidistant bath spins, we assume a uniform distribution
in the interval $\Omega=[a,\,b]$:
\begin{equation}
\rho(z)=\frac{1}{b-a}\;.\label{rho-1-1}
\end{equation}
Finally, from Eq. (\ref{classcont}) and making use of the uniform
density (\ref{rho-1-1}) we obtain for the classical energy density in the thermodynamic limit:

\begin{equation}
\mathcal{E}_{cl}=\lim_{N\to\infty}\,\frac{E_{cl}}{N}=\frac{1}{4\left(a-b\right)}\,\intop_{\Omega}\,\frac{1+k~\sn^{2}(z)}{\sn(z)}\,dz,\label{classcont-1}
\end{equation}

which can be integrated to give
\begin{equation}
\mathcal{E}_{cl}=\frac{1}{4(a-b)}\,\log\left[\frac{\sn(b)\,\left(\cn(a)+\dn(a)\thinspace\right)\,(\dn(b)-k\,\cn(b)\,)}{\sn(a)\,\left(\cn(b)+\dn(b)\thinspace\right)\,(\dn(a)-k\,\cn(a)\,)}\right]\;.\label{Econt-1}
\end{equation}

We will later compare this classical energy density with the large $N$ limit of the exact solution.

\section{Exact solution of the ACSM}

Let us now turn our attention to the exact quantum solution of the
ACSM. We will analyze the distribution of spectral parameters or Bethe
roots $\lambda_{\alpha}$ of the Bethe equations (\ref{eq:Equ}),
as well as the ground state energy (per spin) of the Hamiltoninan
$H_{ACSM}$, for several system sizes up to $N=300$. A large $N$
extrapolation of these results will allow us to compare with the classical
energy density. We choose the modulus $k=0.5$, thus the fundamental
interval is defined by $K(0.5)\approx1.68575$. For the $z$'s interval
we set $a=0.2$ and $b=0.6$.

\begin{figure}
\begin{centering}
\includegraphics[scale=0.5]{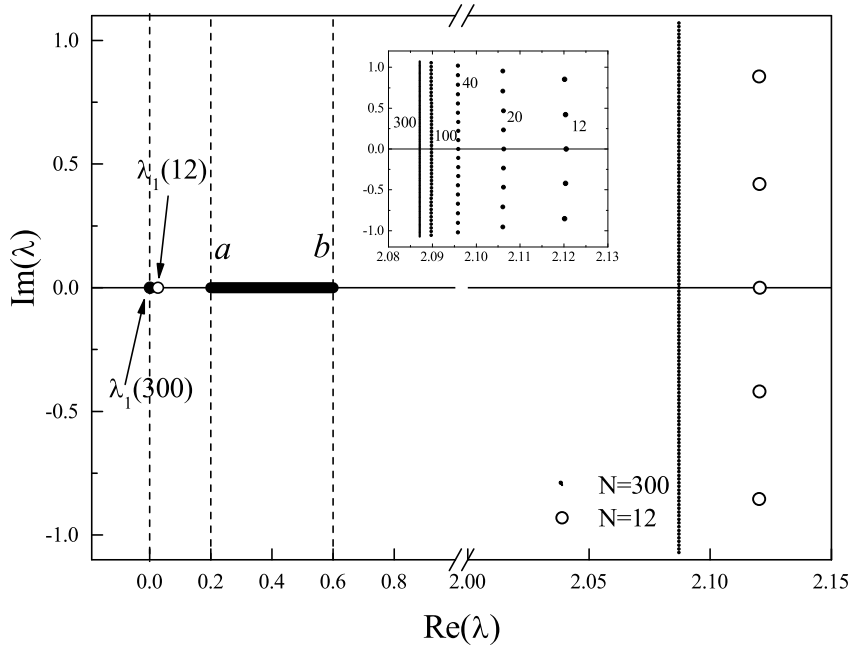}
\par\end{centering}

\centering{}\protect\protect\caption{\label{fig:sol} Ground state Bethe roots of the ACSM (\ref{HCS})
for different system sizes.}
\end{figure}

\begin{table}
\begin{tabular}{|c|c|c|c|c|}
\hline
$N$  & $\lambda_{1}N$  & $\mathrm{Min(Re}(\lambda))$  & $\mathrm{Max(Re}(\lambda))$  & $\frac{E}{N}$\tabularnewline
\hline
12  & 0.327300  & 2.120138  & 2.120487  & -0.821616\tabularnewline
\hline
20  & 0.329466  & 2.106036  & 2.106256  & -0.792253\tabularnewline
\hline
40  & 0.330390  & 2.095787  & 2.095898  & -0.772954\tabularnewline
\hline
80  & 0.330681  & 2.090745  & 2.090800  & -0.7640486\tabularnewline
\hline
100  & 0.330726  & 2.089743  & 2.089787  & -0.762322\tabularnewline
\hline
200  & 0.330805  & 2.087743  & 2.087765  & -0.758920\tabularnewline
\hline
300  & 0.330828  & 2.0870780  & 2.0870926  & -0.757801\tabularnewline
\hline
$\infty$  & 0.330869  & 2.0857505  & 2.0857505  & -0.755586\tabularnewline
\hline
\end{tabular}\protect\caption{Extremum real values of the GS state Bethe roots and corresponding
energies per spin for different $N$ values.}
\label{Extrapola}
\end{table}

The numerical solution of the nonlinear system of equations (\ref{eq:Equ})
faces the usual problems of any Gaudin system, namely a huge number
of independent solutions (the dimension of the system grows exponentially
with $N$) and dangerous divergences which hinders numerical procedures
based on iterative methods. Moreover, finding a specific solution
strongly depends on the initial guess. To overcome these difficulties
we first solve the system for a small number of spins, where we can
identify the distribution of roots for the ground state as well as
for all the excited states. We then follow a particular solution increasing
$N$ by means of an algorithm described below.

In Fig. \ref{fig:sol} we show the distribution of Bethe roots in
the complex plane for the ground states ($l=0\mbox{)}$ of two systems,
a small system with $N=12$ and large system with $N=300$. The complex
plane is divided into three regions delimited by vertical lines crossing
the real axes at $Re(\lambda)=0,\,a,\,b$ and $2K$. We found that
the ground state has the root $\lambda_{1}$, which is associated
with the central spin, always real and located in the first region
satisfying $0<\lambda_{1}<a$, while the other roots are located over
a smooth curve, symmetric with respect to the real axes, with $b<Re(\lambda)<2K$.
We can see in Fig. \ref{fig:sol}, for $N=12$, that the five roots
are almost vertically aligned at $Re(\lambda)\approx2.120$. For $N=300$
the $149$ roots make up an almost vertical segment at $Re(\lambda)\approx2.087$.
The inset amplifies the third region adding several solutions for
the intermediate systems with with $N=20,\,40$ and $100$. For increasing
values of $N$ the real part of the roots in the arc decreases approaching
a limiting value. Moreover, the first root $\lambda_{1}\to0$ in this
limit as can be seen in Table \ref{Extrapola}.

\begin{figure}
\begin{centering}
\includegraphics[scale=0.6]{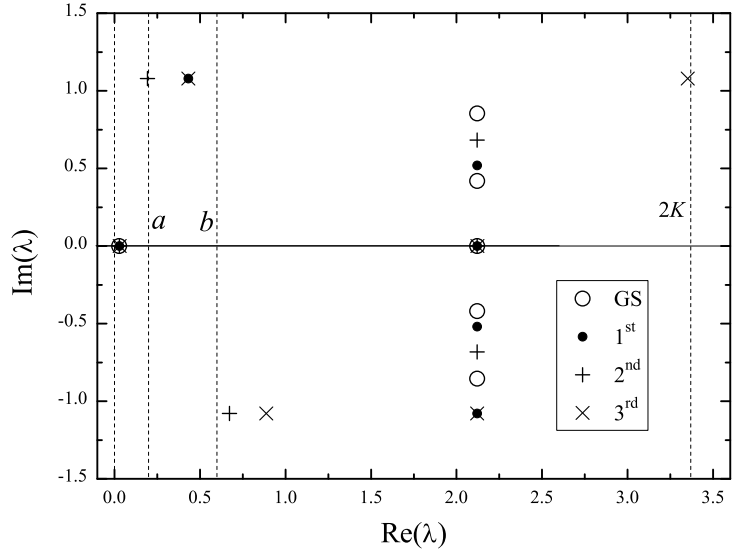}
\par\end{centering}

\centering{}\protect\protect\caption{\label{excited} Bethe roots of the GS and first 3 excited states of the ACSM with N=12 spins.
}
\end{figure}

The ground state always displays the same pattern, with one real root
in the first region $0<\lambda_{1}<a$ and $N-1$ roots distributed
over a smooth arc in the third region, i.e. $Re(\lambda)>b$. A similar
pattern takes place for the lowest energy state in the $l=1$ block.
In Fig. \ref{excited} we show the distribution of Bethe roots for the GS (already displayed in Fig. \ref{fig:sol}) and the first three excited states in positive parity sector of the ACSM with 12 spins. As it can be seen, the arc of complex roots which characterizes the GS persists for the low lying excited states while  some detached roots are distributed in other regions of the complex plane. Different
distributions of the roots give rise to the complete set of eigenstates.

\begin{figure}
\begin{centering}
\includegraphics[scale=0.6]{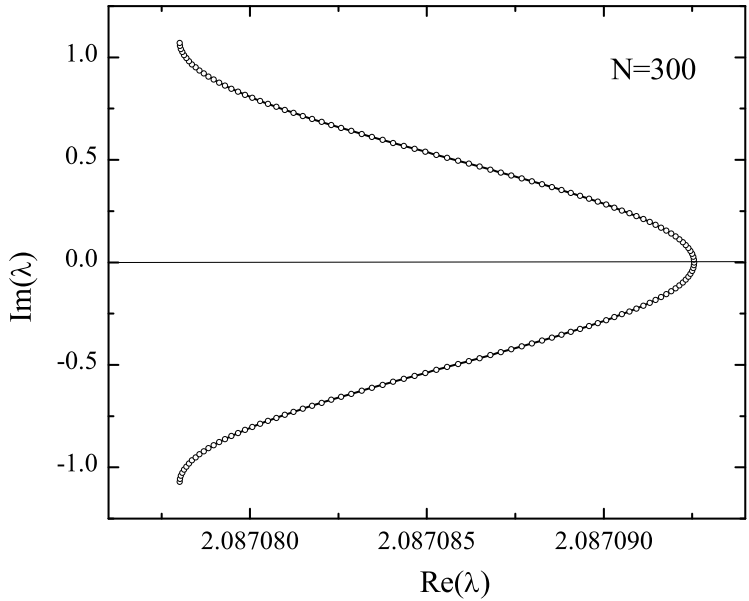}
\par\end{centering}

\centering{}\protect\protect\caption{\label{fig:ArcN300} Ground state Bethe root of the ACSM (\ref{HCS})
for $N=300$. The continuous line corresponds to the fit of Eq. (\ref{FIT}). }
\end{figure}

The scale of the graph in figure \ref{fig:sol}, does not allow to
appreciate the detailed form of the arcs in the third region. In figure \ref{fig:ArcN300} we show, with a smaller scale, the actual
shape of the arc for $N=300$. The difference between the maximum
and minimum real parts of the roots in the arc decreases with increasing
$N$, becoming null at $N\to\infty$. Therefore, in the continuous
limit the arc turns into a vertical segment with $\re(\lambda)=2.08575$
with half of the quasi-periods as the interval extremes $-\frac{K'}{2}\leq\im\left(\lambda\right)\leq-\frac{K'}{2}$
(see table \ref{Extrapola}).

In order to obtain the numerical ground state solution for a large
$N$ system, we start from the solution of a small system. We then
increase $N$, typically doubling it in each iteration. In each step
we make a least square fit of the complex arc expressing the real
part $\re(\lambda)=x$ as a function of the imaginary part $\im(\lambda)=y$
taken as the independent variable. An excellent agreement is obtained
for any $N$ value by means of the 4-parameter function:

\begin{equation}
x=\alpha+\beta\,\dn(c_{1}\,y)\,\cn(c_{2}\,y)\;.\label{FIT}
\end{equation}

The continuous line in Fig. \ref{fig:ArcN300} corresponds to a fit
of this function for the $N=300$ system. We take advantage of this
excellent fit to generate, for each system of size $N$ the initial
guess from the lower size system $N_{0}$. In addition, the initial
guess for the first root $\lambda_{1}$ is obtained by a linear scaling
$\lambda_{1}=\lambda_{1}^{0}\frac{N_{0}}{N}$. In both cases the index
$0$ stands for the solution of the previous step. The procedure is
stable, and it allows to find numerical solutions for very large systems.

\begin{figure}
\begin{centering}
\includegraphics[scale=0.4]{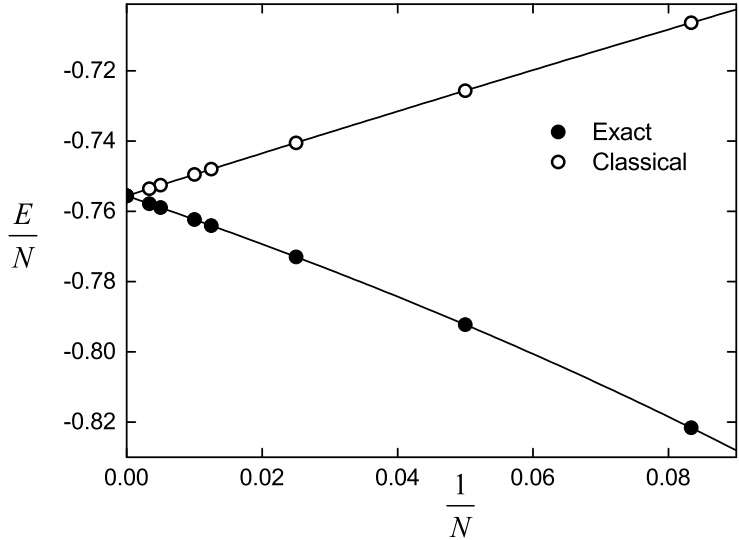}
\par\end{centering}

\centering{}\protect\protect\caption{\label{classener}Exact quantum and classical GS energies per spin
of the ACSM for different $N$ values. The extrapolated $N\to\infty$ limits coincide with the analytic value (\ref{Econt-1}).
}
\end{figure}

It is interesting to analyze the numerical results in the $N\to\infty$
limit. Table \ref{Extrapola} shows these results for several values
of $N$ and the numerical extrapolation to the thermodynamic limit.
The fifth column of the table displays the energy per site and the
extrapolated value in the thermodynamic limit. We can now compare
these results with the classical energy density. For $a=0.2$, $b=0.6$ and
$k=0.5$ equation (\ref{Econt-1}) yields $\mathcal{E}_{cl}=-0.75558603$
showing an excellent agreement with the extrapolated result. In figure
(\ref{classener}) we show a comparison between the classical energy
(\ref{enerclas2-1}) (open circles) and the quantum energy (black
circles) for several values of $N$. Both branches converge to a common
limit for $N\to\infty$. Continuous lines correspond to a third order
polynomial least square fits. The gap between the classical and the
quantum energies for finite systems is a direct consequence of the
quantum fluctuations that disappear in the thermodynamic limit.

\section{Summary}

The rational and hyperbolic Gaudin models have been extensively employed
lately to study many-body quantum systems in several branches of mesoscopic
physics. On the contrary, the EGM derived by Gaudin in 1976 and solved
exactly with the ABA in 1996 has been scarcely used as a mathematical
tool to investigate many-body physical problems. In this article we
have first summarized the integrals of motion of the model and the
exact solution for systems with arbitrary spins. Subsequently, we
discussed the behavior of the Bethe roots for the complete set of
eigenstates of a system with three different spins. We showed that
the Bethe roots should be restricted to the fundamental rectangle
in order to warrant that every solution corresponds to a physical
eigenstate. Finally, we introduced a particular anisotropic CSM, accounting
for the hyperfine interaction of an electronic spin in a
quantum dot with the environmental nuclear spins, and possible effects due to quadrupole couplings. The so called ACSM
was solved exactly for large number of spins and the GS energy
was extrapolated to the thermodynamic limit, which coincides with
the classical approximation. We hope that our numerical studies would
pave the way to applications of the EGM to other physical many-body
systems.

\begin{acknowledgements} This work was supported by grant FIS2012-34479
of the Spanish Ministry of Economy and Competitiveness. \end{acknowledgements}

\section*{Appendix. Generalization of the EGM to arbitrary spins}

We start with the known integrals of motion, the corresponding
eigenvalues and the Bethe equations for spin $1/2$ systems \cite{babujian}

\begin{equation}
H_{a}=\sum_{b\left(  \neq a\right)  =1}^{\mathcal{N}}\left[  J^{x}\left(
z_{a}-z_{b}\right)  ~S_{a}^{x}S_{b}^{x}+J^{y}\left(  z_{a}-z_{b}\right)
~S_{a}^{y}S_{b}^{y}+J^{z}\left(  z_{a}-z_{b}\right)  ~S_{a}^{z}S_{b}%
^{z}\right],
\end{equation}

\begin{equation}
h_{a}=\frac{1}{4}\sum_{b\left(  \neq a\right)  =1}^{\mathcal{N}}\varphi\left(
z_{a}-z_{b}\right)  -\frac{1}{2}\sum_{\alpha=1}^{\mathcal{M}}\varphi\left(
z_{a}-\lambda_{\alpha}\right)  +\frac{i\pi l}{4K},
\end{equation}

\begin{equation}
\frac{1}{2}\sum_{a=1}^{\mathcal{N}}\varphi\left(  \lambda_{\alpha}%
-z_{a}\right)  -\sum_{\beta\left(  \neq\alpha\right)  =1}^{\mathcal{M}}%
\varphi\left(  \lambda_{\alpha}-\lambda_{\beta}\right)  +\frac{i\pi l}{2K}=0,
\end{equation}
where $\mathcal{N}$ is the number of spins and $\mathcal{M}=\mathcal{N}/2$ the
number of roots.

Following Ref. \cite{Gaudin} we now group an arbitrary number of adjacent spins into clusters
and define a new lattice with sites $i$ and spins $S_{i}=\sum_{a=a_{i}}%
^{a_{i}+n_{i}}S_{a}$, where $a_{i}$ is leftmost site of the cluster $i$ and
$n_{i}$ the number of spins in the cluster. The new lattice fulfills
$\mathcal{N}=\sum_{i=1}^{N}n_{i}$, where $N$ is the number of clusters or
sites in the new lattice. We recall that the functions $J\left(  x\right)  $
and $\varphi\left(  x\right)  $ are odd functions. The new integrals of motion are%

\begin{align*}
R_{i} &  =\sum_{a=a_{i}}^{a_{i}+n_{i}}H_{a}=\sum_{a=a_{i}}^{a_{i}+n_{i}}%
\sum_{b\left(  \neq a\right)  =a_{i}}^{a_{i}+n_{i}}\left[  J^{x}\left(
z_{a}-z_{b}\right)  ~S_{a}^{x}S_{b}^{x}+J^{y}\left(  z_{a}-z_{b}\right)
~S_{a}^{y}S_{b}^{y}+J^{z}\left(  z_{a}-z_{b}\right)  ~S_{a}^{z}S_{b}%
^{z}\right]  +\\
&  \sum_{a=a_{i}}^{a_{i}+n_{i}}\sum_{b\notin C_{i}}\left[  J^{x}\left(
z_{a}-z_{b}\right)  ~S_{a}^{x}S_{b}^{x}+J^{y}\left(  z_{a}-z_{b}\right)
~S_{a}^{y}S_{b}^{y}+J^{z}\left(  z_{a}-z_{b}\right)  ~S_{a}^{z}S_{b}%
^{z}\right]  \,\text{\ },
\end{align*}
where $C_{i}\equiv\left\{  a_{i},a_{i+1},\cdots,a_{i}+n_{i}\right\}  $. \ The
first term in the right hand side cancels out due to the antisymmetry of the
functions $J\left(  x\right)  $. We now assume that the parameters $z$ inside
each cluster are all equal, $z_{a}=z_{i}$ for all $a\in C_{i}$, and together
with the definition of the cluster spins $S_{i}$ we obtain the integrals of
motion in the general case %

\begin{equation}
R_{i}=\sum_{j\left(  \neq i\right)  =1}^{N}\left[  J^{x}\left(  z_{i}%
-z_{j}\right)  ~S_{i}^{x}S_{j}^{x}+J^{y}\left(  z_{i}-z_{j}\right)  ~S_{i}%
^{y}S_{j}^{y}+J^{z}\left(  z_{i}-z_{j}\right)  ~S_{i}^{z}S_{j}^{z}\right].
\end{equation}

The corresponding eigenvalues and Bethe equations are transformed as%

\begin{equation}
r_{i}=\sum_{a=a_{i}}^{a_{i}+n_{i}}h_{a}=\frac{n_{i}}{4}\sum_{j\left(  \neq
i\right)  =1}^{N}n_{j}\varphi\left(  z_{i}-z_{j}\right)  -\frac{n_{i}}{2}%
\sum_{\alpha=1}^{\mathcal{M}}\varphi\left(  z_{i}-\lambda_{\alpha}\right)
+i\frac{\pi l}{4K}n_{i}%
\end{equation}
and%

\begin{equation}
\frac{1}{2}\sum_{i=1}^{N}n_{i}\varphi\left(  \lambda_{\alpha}-z_{i}\right)
-\sum_{\beta\left(  \neq\alpha\right)  =1}^{\mathcal{M}}\varphi\left(
\lambda_{\alpha}-\lambda_{\beta}\right)  +i\frac{\pi l}{2K}=0 .
\end{equation}

It can be shown \cite{Gaudin} that the exact solution corresponds to the maximum
spin  in each cluster, namely $s_{i}=\frac{n_{i}}{2}$, which implies that
$M=\mathcal{M=}\frac{1}{2}\sum_{i}n_{i}=\sum_{i}s_{i}$. With these last
replacements we obtain the eigenvalues (\ref{autov}) and Bethe equations (\ref{eq:Equ}).

\end{document}